\documentstyle[aps,epsf]{revtex}  

\begin{document}        

\baselineskip 14pt
\title{Particle Identification with BELLE}
\author{Asish Satpathy}
\address{University of Cincinnati, Ohio, OH 45221, USA }
%
\maketitle              

\begin{abstract}        

The working principle and performance of the BELLE particle identification 
device (PID), based on a hybrid system consisting of an array of 
high precision scintillator Time of Flight and silica Aerogel Counters, 
is discussed. The performances achieved in the beam tests are satisfactory
and Monte Carlo tests of meeting physics objectives of BELLE
are promising. Prior to the real experiment which is expected to 
commence in spring 1999, the BELLE PID is taking cosmic ray data for 
calibration and fine tuning.

\end{abstract}   	

\section{Introduction}               

The new KEK $B$ asymmetric $e^{+} e{-}$ collider 
is scheduled to be fully operational in spring 1999.
The BELLE experiment at KEK $B$ is in a preparatory stage of data taking
and aims at either confirm or refute the present 
hypothesis of the Standard Model of 
CP violation in terms of phases in the Cabibbo-Kobayashi-Maskawa 
(CKM) matrix through a varieties of CP asymmetries in neutral and
charged $B$ decays \cite{TDR}.
One of the most important task of the 
BELLE detector is the identification of the charged 
hadrons which is relevant for the reconstruction 
of many beauty and charm decay channels.
This facilitates (1) the $B$ flavor tagging
which relies on the correlation between the charge of kaon and the 
flavor of the decaying $B^{0}$ from which it is originated, and (2) 
identification of exclusive final states such as 
$B^{0} \rightarrow \pi^{+} \pi^{-}$ ,
that will provide a measurement of the angle 
$\alpha$ in the unitary triangle. Since the relative abundance
of pions and kaons in $B$ decays are approximately 8:1, the $B$ 
flavor tagging requires good kaon identification with minimal
pion contamination. Also the $b \rightarrow u$ type suppressed
mode requires good separation from the penguin type 
decay such as $B^{0} \rightarrow K^{+} \pi^{-}$ 
which is of a similar magnitude. 

Physics requirements divide the momentum coverage of charged hadron
identification into two approximately non overlapping regions.
Flavor tagging of the opposite $B^{0}$ though the detection 
of charged kaons produced in $b \rightarrow c \rightarrow s$ cascade
requires kaon identification in the low momentum region i.e.
between 0.2 GeV/c and 1.5 GeV/c (Fig 1). Due to the boost of the 
$B \bar{B}$ system the $\pi$ momenta range from about 1.5 GeV/c
to 4.0 GeV/c for two body decays such as $B^{0} \rightarrow \pi^{+} \pi^{-}$
(Fig 2.)

\begin{figure}[hb]	
\centerline{
\epsfxsize 2.0 truein \epsfbox{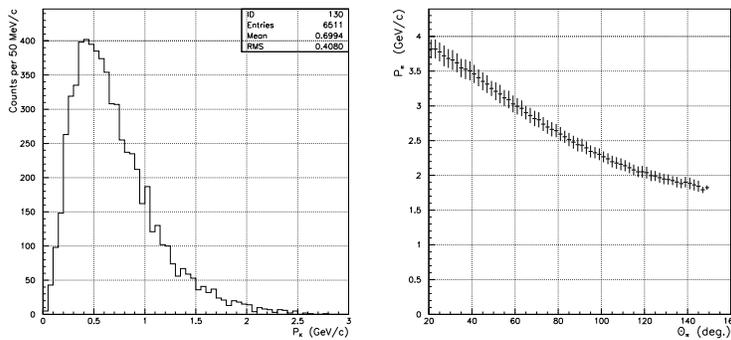}   
\epsfxsize 2.0 truein \epsfbox{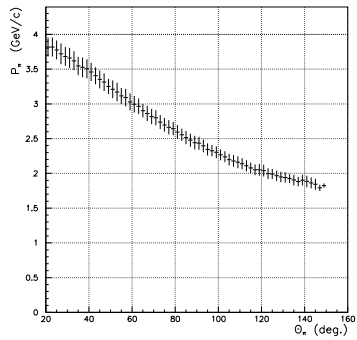}}   
\vskip 0.0 cm
\caption[]{ (Left) Momentum spectrum of charged kaons arising from the 
$b \rightarrow c \rightarrow s $ cascade.}
\caption[]{ (Right) Momentum of pions from 
$B^{0} \rightarrow \pi^{+} \pi^{-}$ as a function of the laboratory 
polar angle. }

\label{LegsFigure}
\end{figure}

In addition to the above requirement, the particle identification (PID)
system should have a minimum inactive material in front of the CsI(Tl)
crystal calorimeter to preserve the good energy resolution and detection 
efficiency of soft photons. A sufficient sensitivity of signal to
noise over the full angular and momentum range would be an added
feature and the system must be able to operate efficiently in a 1.5 T magnetic
field. \\

Considering the above requirements,
a hybrid system consisting of an array of scintillator Time
of Flight (ToF) counters and an array of silica Aerogel Cherenkov Counter (ACC)
has been chosen as the BELLE particle identification device where
the ToF counters cover the momentum region below 1.2 GeV/c and ACC
provide identification at higher momenta. This approach has an advantage
of simplicity.

\section{The Time of Flight Counters }
The subsystem consists of 128 Bicron BC408 Scintillator counters
and 64 BC412 Trigger Scintillation (TSC) counters with fine mesh
photo multiplier tubes (FMPMT) to read out the signals. 
By using the FMPMTs one eliminates the need for
the light guides which gives a big reduction of time resolution.
These modules are individually mounted to the inner wall of the CsI container 
at 1175 cm radius from the beam axis. The angular
coverage of ToF is $33.7 ^{\circ} < \theta < 120.8 ^{\circ} $.
They are used to start 
a clock and stop counting at a precise time ( with less than
100 psec time resolution ) after a beam crossing 
take place, thereby allowing the determination of the time 
it takes a particle to travel from the center of the interaction
point to the ToF layer. This time, together with the knowledge
of particle's momentum from Central Drift Chamber (CDC), 
allows an estimate of particle mass and thus identity.
\vspace*{0.5cm}
\begin{figure}[hb]	
\centerline{
\epsfxsize 3.0 truein \epsfbox{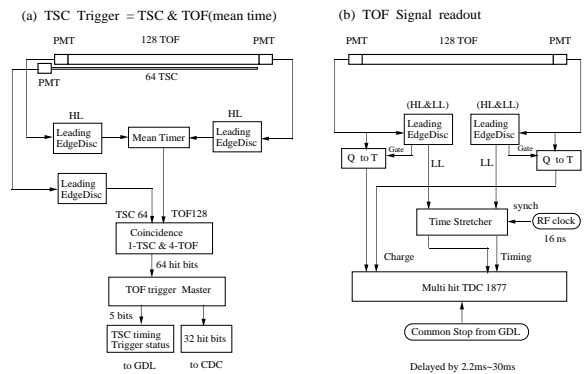} }  
\caption[]{Block diagrams of the ToF electronics for (a) trigger
and (b) readout.}
\label{LegsFigure}
\end{figure}

In addition to the $B$-flavor tagging capability,
ToF provides an event timing signal used by the trigger
to provide a gate to the readout of 
other sub-detectors such as Electromagnetic Calorimeter (ECL) and CDC.
Fig 3 shows the schematic diagram for the (a) fast trigger
and (b) ToF readout. 
The FMPMT signals are read out into 
fast leading edge discriminators. A high level threshold 
discriminator is used to gate the low level timing
signal. The signals are also read into MQT300A Charge to Time conversion 
chips. The BELLE standard readout board, LeCroy 1877s multi-hit TDC's,
are used to read out the signals from the MQT300A chip. These TDC's
have a least significant bit (LSB) of 500 ps.
With the collaboration with LeCroy a "Time Stretcher" modules
have been developed \cite{gary}. 
The time stretcher expands a 25 ps LSB into 
a 500 ps LSB. By reading the timing information into the time stretcher and
1877s combination and 
performing the time walk correction, we have achieved 25ps  
resolution in the readout electronics which gives us a 100 ps overall
timing resolution.

The forward and backward FMPMT timing signals of ToF modules are mean timed.
The TSC timing signal is used to gate this mean time signal.
The first mean time in the event is used for an on line event timing
signal for the CDC and for fast reconstruction on the on-line farm.
Xilinx pipelines are also used to calculate the event multiplicity and
event shape (in $\phi$) for background reduction in the trigger. 
\subsection{Results from Beam Tests}
Full size prototype ToF counters  (BC408 scintillators 
with attenuation length about 2.5 m ) were tested using a $\pi$ beam
by placing the counter on a movable stage that could be rotated
around a pivot point \cite{prog}. 
Fig 4 shows the time resolution 
as a function of beam position z. A time walk correction has been
applied over all beam position and the time jitter of the start 
counter ($\sim$ 35 ps) was subtracted quadratically. 
Intrinsic time resolution of 85 ps are obtained with a discriminator
threshold set at 100 mV. Fig 5 shows the $\pi^{+}$/$p$ separation
for a 2 GeV/c un-separated beam. The observed 6$\sigma$ separation 
between $\pi^{+}$ and $p$ corresponds to what could be expected
for $\pi$/$K$ separation in a 1 GeV/c beam. The separation is improved
near the FMPMT owing to the longer path length and better timing
resolution. 

\vspace*{0.0cm}
\begin{figure}[hb]	
\centerline{
\epsfxsize 3.0 truein \epsfbox{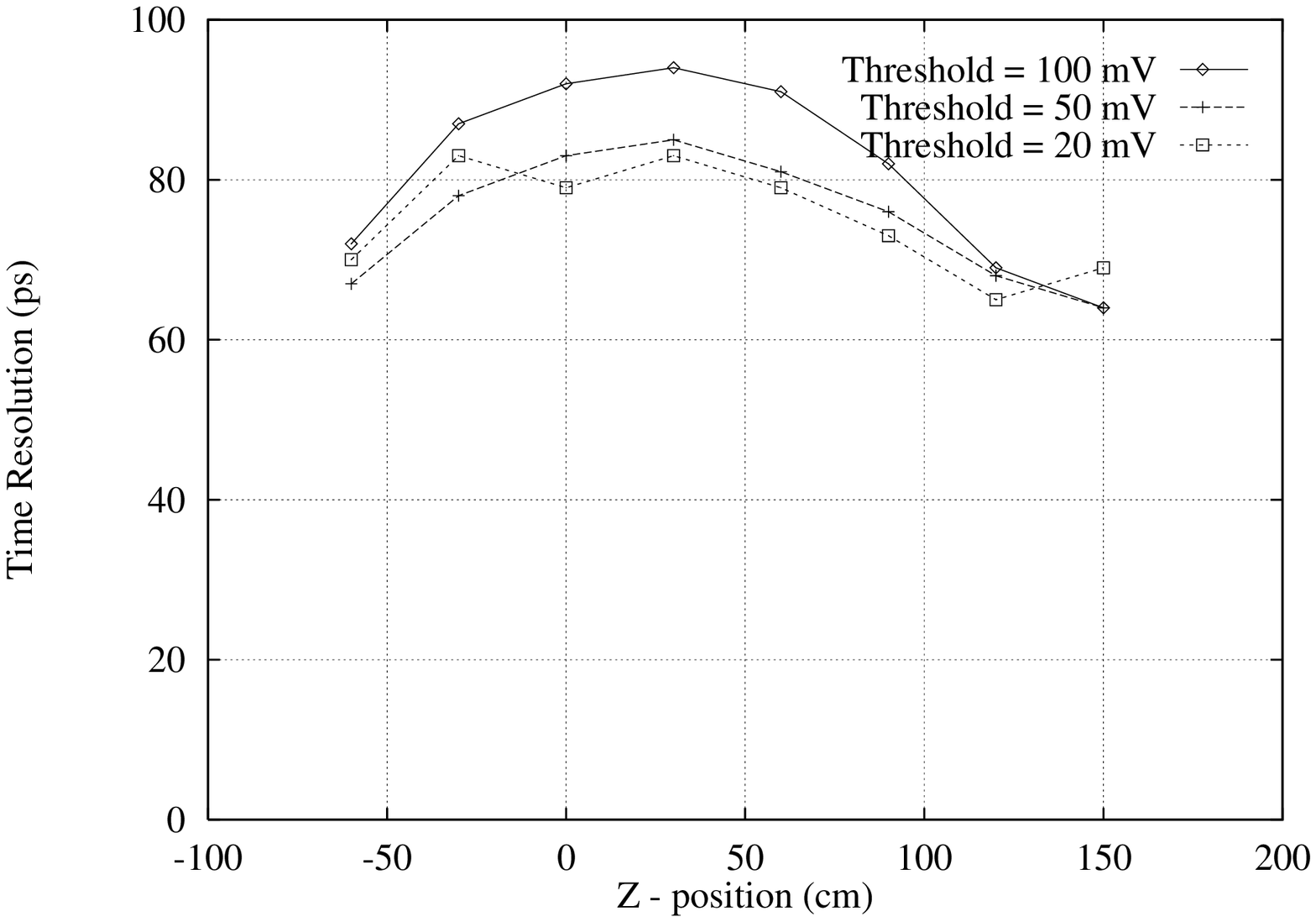}
\epsfxsize 1.7 truein \epsfbox{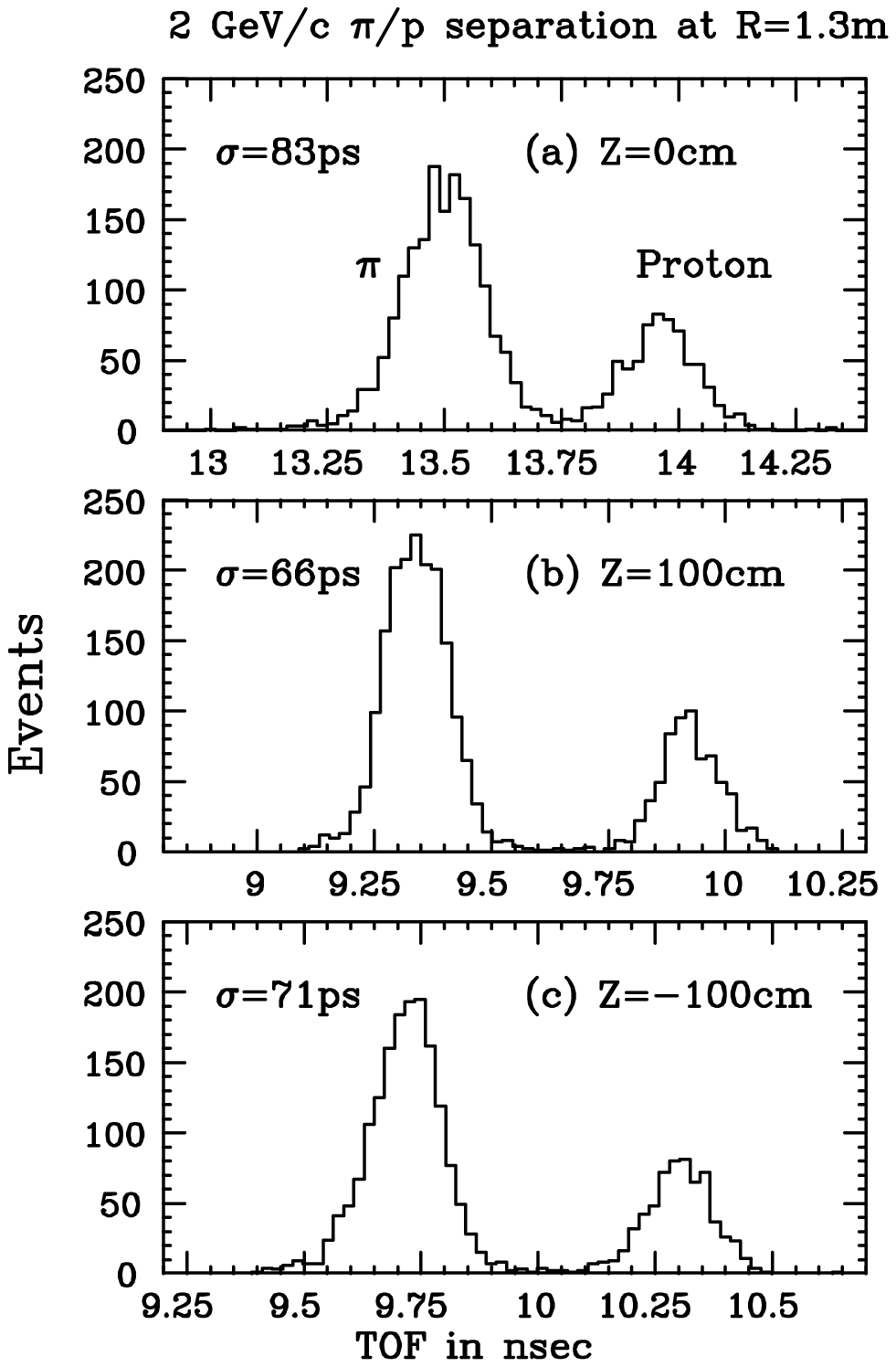}}   
\vskip 0.4 cm
\caption[]{ (Left) Time resolution vs. Z for a full size prototype of 
BELLE ToF.}
\caption[]{ (Right) $\pi$/proton separation in a 2 GeV/c beam which contains
both $\pi$ and protons.}
\label{LegsFigure}
\end{figure}

\section{ The Aerogel Cherenkov Counters  }
The ACC sub-system consists of 960 element-arrays (16 elements in 
$z$ and 60 elements in $\phi$) in the barrel and 268 element-array in 
the forward endcap.
The barrel ACC (BACC) system is located between the CDC and CsI starting at 
an inner radius of 88.5 cm with the z coverage of  
-85 $< z <$ 162 cm. The forward endcap ACC (EACC) is located between
the forward endcap CsI and CDC endplate occupying the region
bounded by 42$<r<$114 cm and 116$<z<$194 cm (Fig 7). 
Fig 6 shows the schematic of a single aerogel block in the array.
Each aerogel block is made up of 5 
layers of silica aerogel slabs housed in light-weight and 
light tight aluminum boxes.
Being a Rayleigh scatterer, mean path lengths in aerogel are larger than
in any non-scattering media. Therefore absorption in the aerogel and 
on the container walls is minimized by covering the wall with highly 
efficient white diffuse reflector (Gortex Teflon). Since the
detector stays within a 1.5 T magnetic field, FMPMT 
are used to detect Cherenkov radiation. 
With a proper choice of refractive index, 
charged pions cause light to be emitted in the aerogel.
The refractive indices for BACC varies with $\theta$
(n = 1.01, 1.013, 1.015, 1.020, 1.028) 
in a way that takes into account the general
softening of the hadron momentum spectrum with increasing lab
polar angle. All the EACC counters use n=1.03 aerogel that
are appropriate for flavor tagging in the momentum region between
0.8 and 2.5 GeV/c.
Table 1 summarizes the $\pi$, $K$ and $p$ thresholds for 
different refractive indices used in the design. Around 3.5 GeV/c,
$K$ also produces Cherenkov light and so separation of $K$ and
$\pi$ becomes difficult in the forward direction when the 
particle has momentum more than 3.5 GeV/c. One can somehow get
around the problem by looking at the pulse height information
of the Cherenkov light since $\pi$ mesons tend to have larger pulse
heights.
\begin{figure}[hb]	
\centerline{
\epsfxsize 2.5 truein \epsfbox{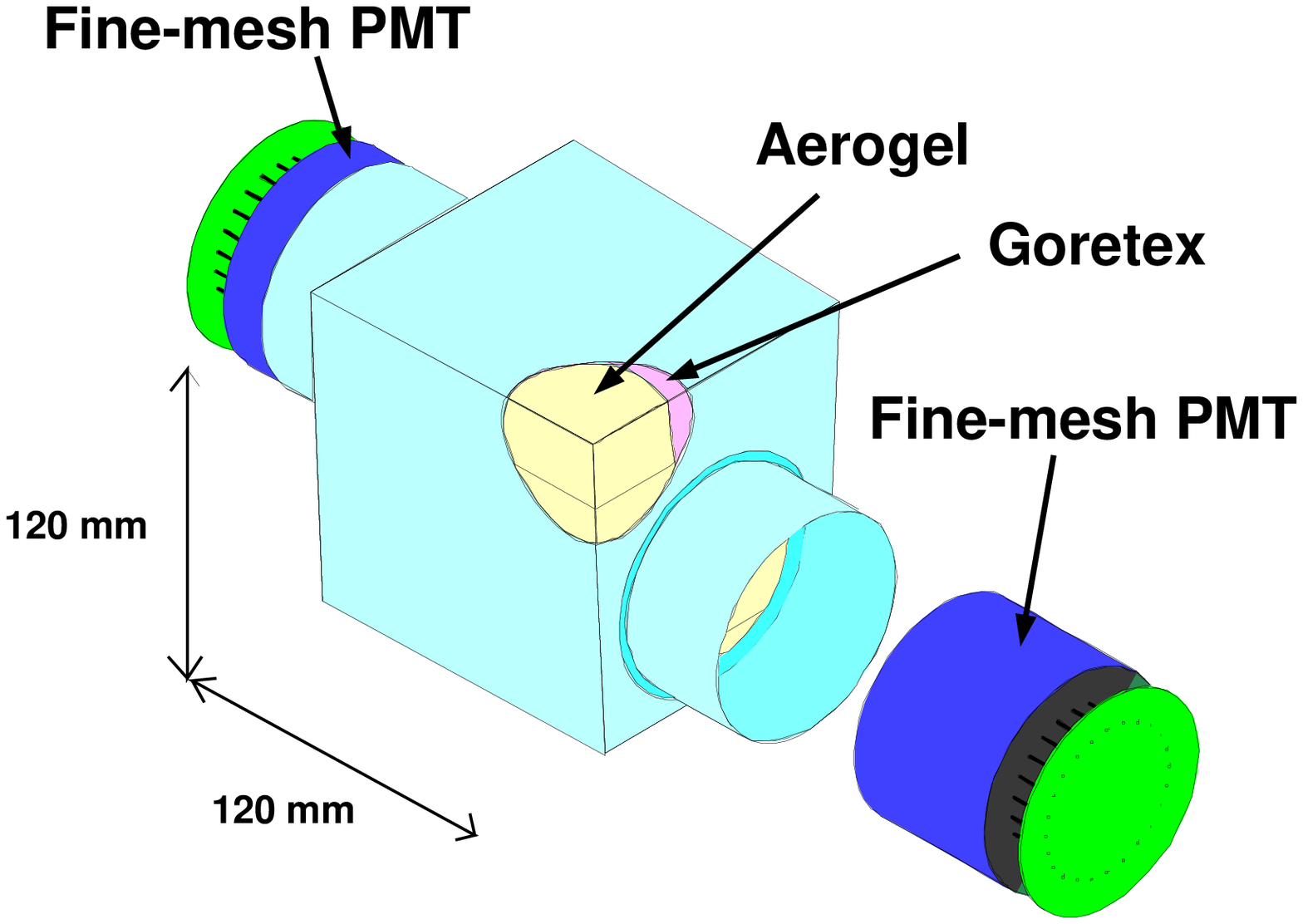}   
\epsfxsize 3.5 truein \epsfbox{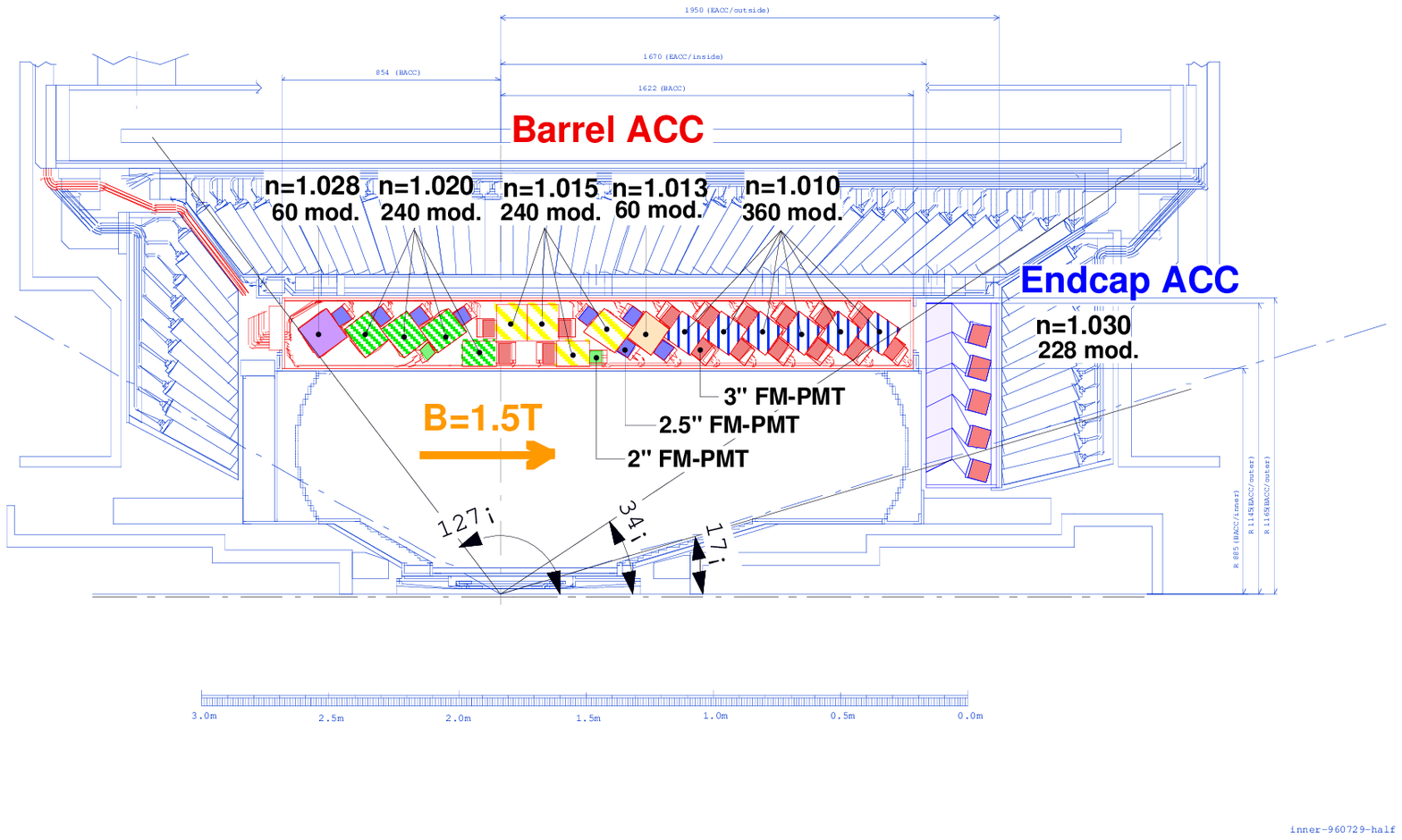}}
\vskip 0.0 cm
\caption[]{ (Left) Configuration of single ACC Module. }
\caption[]{ (Right) Side view of the arrangement of the barrel and 
forward endcap Cherenkov Counters. }
\label{LegsFigure}
\end{figure}

\begin{table}
\begin{center}
\caption{Momentum thresholds and Cherenkov angles for
radiators of different refractive indices .}
\begin{tabular}{c c c c c} 
n & $p_{\pi}$ [GeV/c]  & $p_{K}$ [GeV/c]  & $p_{p}$ [GeV/c] & $\theta_{c}$ \\
\tableline \tableline
1.010 &0.985 &3.482 &6.618 &$8.069^{\circ}$ \\ 
1.013 &0.863 &3.052 &5.800 &$9.189^{\circ}$ \\
1.015 &0.803 &2.840 &5.397 &$9.862^{\circ}$ \\
1.020 &0.695 &2.456 &4.668 &$11.365^{\circ}$ \\
1.028 &0.585 &2.072 &3.937 &$13.403^{\circ}$ \\
1.030 &0.566 &2.000 &3.802 &$13.862^{\circ}$ \\
\end{tabular}
\end{center}
\end{table}
Besides having low refractive indices, the aerogels are hydrophobic
- that ensures long term stability of the detector. In a separate
test, aerogels were found to be radiation hard up to 
10 MRad equivalent dose \cite{sahu}.

Since the gain of FMPMT drops sharply in high magnetic fields,
one needs further amplification of the Cherenkov signal. Depending on the
threshold, PMT signal from either $\pi$'s or $K$'s is amplified about
10 times before it goes to MQT300 chips.
The output from the chips is fed to the Lecroy's 1877s TDCs whose leading edge
gives the timing of the pulse and the width is proportional to the 
pulse amplitude.
\subsection{Test Beam Results}
\vspace*{0.0cm}
\begin{figure}[h]	
\centerline{
\epsfxsize 3.5 truein \epsfbox{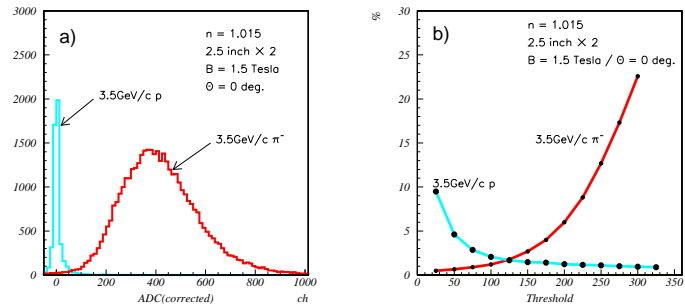}}  
\vskip 0.5 cm
\caption[]{ (a) The pulse height distribution obtained with 3.5 GeV/c pions in a magnetic field for n=1.5 T (b) The inefficiency and background contamination as a function of pulse height threshold values.}
\label{LegsFigure}
\end{figure}
Fig 8 demonstrates the performance of ACC prototype achieved during 
a test beam operation with 1.5 Tesla magnetic field \cite{prog}. 
A pulse height 
distribution separating two 3.5 GeV/c charged particles, 
pions and protons,
with 
\newpage
n = 1.05 aerogel is shown in Fig 8(a).
More than 4 $\sigma$ separation with efficiency better 
than 98 \% can be achieved with less than 
2 \% background contamination which is mainly coming from proton induced 
knock-on electrons produced  on the 1 mm thick aluminum box and aerogel 
material itself. It should be noted that in practice, unlike the beam
test, the fake rate is dominated by hits on the glass window of
FMPMT and showering in FMPMTs \cite{suda}. 
Fig 8(b) shows the efficiency and background 
contamination as a function of threshold on the pulse height.
The average number of photoelectrons obtained for 3.5 GeV/c pions incident 
at the center of the counter was found to be 20.3 when viewed by 
two 2.5 inch FMPMTs. Considering the above results
and the momentum dependence of Cherenkov light yield, we expect that 
more than 3 $\sigma$ $\pi/K$ separation is possible in the momentum region
of 1.2 to 3.5 GeV/c for BACC and 0.8 to 2.2 GeV/c for the 
EACC.
\section{Simulation Results}
To demonstrate that designed PID fulfills the various
physics requirements, a number of simulation studies (both parametrized and
detector based) has been done. Figure 8 shows an example of 
good separation between signal events
$B^{+} \rightarrow \bar{D^{0}}K^{+}$ when (a) $\bar{D^{0}} 
\rightarrow K^{+} \pi^{-}$
(b) $\bar{D^{0}} 
\rightarrow K^{+} \pi^{-} \pi^{+} \pi^{-}$ and backgrounds coming 
from the mis-identification of kaons, after the data are normalized 
to the same integrated luminosity.
\begin{figure}[hb]	
\centerline{
\epsfxsize 1.9 truein \epsfbox{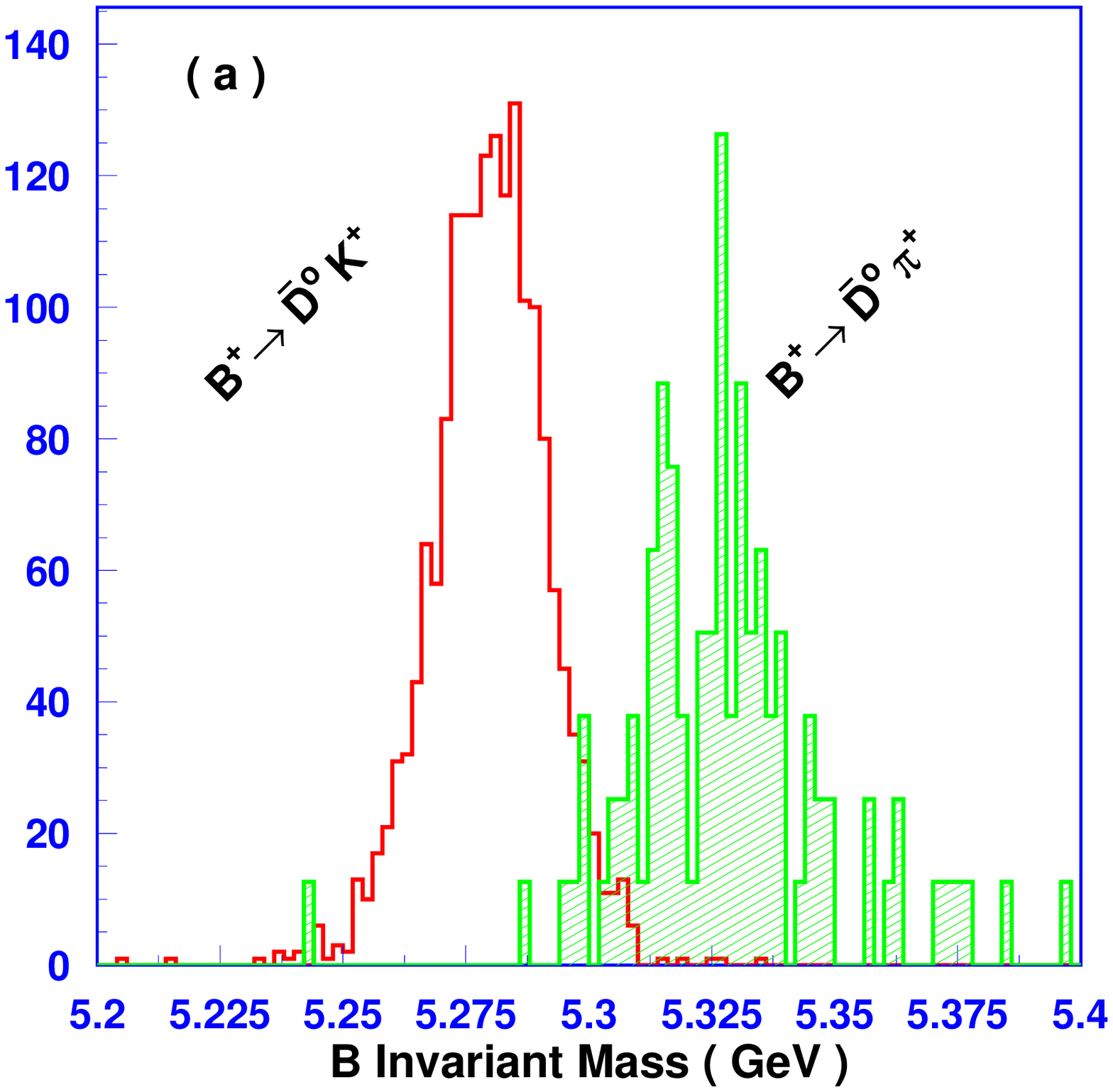}   
\epsfxsize 1.9 truein \epsfbox{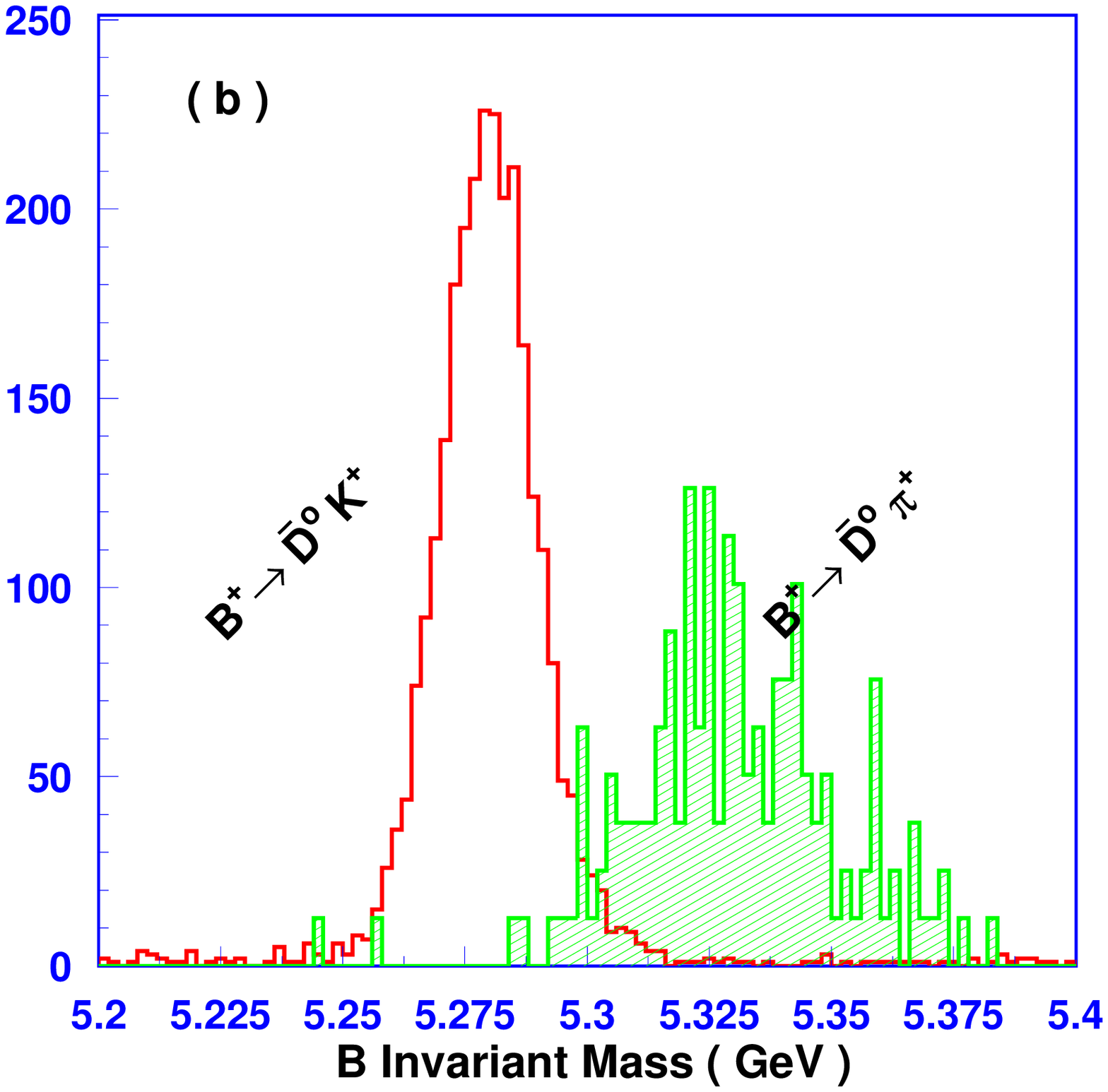}}
\vskip 0.0 cm
\caption[]{Demonstration of PID performance in separating signal and 
background events in $B$ decays.}
\label{LegsFigure}
\end{figure}
Similar simulation studies have been done for other $B$ decay modes that 
contain $\pi$'s and $K$'s in the decay products 
and the results are found satisfactory.
\section{Conclusion}
A PID system based on hybrid system of ToF and silica aerogel counters
is simple, robust and a powerful device that provides excellent particle
identification over entire solid angle and momentum range. 
The detector has been constructed at KEK and being calibrated and tuned with
cosmic ray events at a roll out position 
before the data taking with beam starts in spring 1999.

\end{document}